\documentclass[11pt,a4paper]{article}
\pdfoutput=1
\usepackage{jheppub}

\usepackage{amsfonts, bm, slashed}

\begin{document}

\begin{flushright}
KCL-PH-TH/2015-38\\ TUM-HEP-1010-15
\end{flushright}

\title{Non-Hermitian extension of gauge theories
and implications for neutrino physics}

\author[a]{Jean~Alexandre,}
\author[a,b]{Carl~M.~Bender,}
\author[c]{and Peter~Millington}

\affiliation[a]{Department of Physics, King's College London,
Strand WC2R 2LS, London, United Kingdom}
\affiliation[b]{Department of Physics, Washington University,
One Brookings Drive, St.~Louis, MO 63130, USA}
\affiliation[c]{Physik Department T70, Technische Universit\"{a}t M\"{u}nchen, James-Franck-Stra\ss e,
85748 Garching, Germany}

\emailAdd{jean.alexandre@kcl.ac.uk}
\emailAdd{cmb@wustl.edu}
\emailAdd{p.w.millington@tum.de}

\abstract{
An extension of QED is considered in which the Dirac fermion has both Hermitian
and anti-Hermitian mass terms, as well as both vector and axial-vector couplings
to the gauge field. Gauge invariance is restored when the Hermitian and
anti-Hermitian masses are of equal magnitude, and the theory reduces to that of
a single massless Weyl fermion. An analogous non-Hermitian Yukawa theory is
considered, and it is shown that this model can explain the smallness of the
light-neutrino masses and provide an additional source of leptonic ${\cal CP}$ violation.
}

\arxivnumber{1509.01203}

\maketitle
 
\section{Introduction}
\label{s1}

Discrete symmetries play a fundamental role in particle physics. 
Charge conjugation (${\cal C}$) and the discrete spacetime symmetries of parity (${\cal P}$) 
and time reversal (${\cal T}$) are such that ${\cal CPT}$ is
necessarily conserved for a local, Lorentz-symmetric, and Hermitian theory. 
There is, however, no reason 
for all of the latter requirements to be essential in the 
building of viable models. This has been
shown, for example, in ref.~\cite{Barenboim}, where locality is dropped, leading
to a Lorentz-symmetric description of neutrino physics, which is \emph{odd} under 
${\cal CPT}$. In the present article we keep locality but consider a 
non-Hermitian (Lorentz-symmetric) model.
 
The last 15 years have seen much interest in and research activity on theories
described by non-Hermitian Hamiltonians. Such theories have remarkable and often
unexpected properties. For example, the eigenvalues of the non-Hermitian
${\cal CPT}$-symmetric quantum-mechanical Hamiltonians $H=p^2+ix^3$ and $H=p^2-x^4$
are real, positive, and discrete \cite{R1,R2}.

One idea that has been pursued repeatedly is to study the properties of a
non-Hermitian version of quantum electrodynamics (QED). The Hamiltonian for QED
becomes non-Hermitian if the unrenormalized electric charge $e$ is chosen to be
imaginary. Then, if the electric potential is chosen to transform as a
pseudovector rather than a vector, the Hamiltonian becomes ${\cal CPT}$ symmetric. The
resulting non-Hermitian theory of electrodynamics becomes a multi-component
analog of a self-interacting spinless quantum field theory (QFT), comprising a
pseudoscalar field $\phi$ with a cubic self-interaction term of the form $i
\phi^3$. This pseudoscalar QFT was studied in detail in ref.~\cite{R3}, and the
non-Hermitian version of QED was studied in ref.~\cite{R4}. This non-Hermitian
version of electrodynamics is particularly interesting because it is
asymptotically free and the version of this theory with massless fermions
appears to have a nontrivial fixed point (see refs.~\cite{R5a,R5b,R5c}). A perturbative
calculation of a metric with respect to which this theory is unitary is given in
ref.~\cite{R6}.

A detailed analysis of ${\cal CPT}$-symmetric non-Hermitian fermionic theories was
done by Jones-Smith and Mathur~\cite{R7}. In this work it was emphasized that 
for fermions the time-reversal operator ${\cal T}$ has the property that ${\cal T}^2=
-1$. This represents a significant departure from the case of bosonic theories, 
where ${\cal T}^2=1$. (Further work on the properties of ${\cal CPT}$-symmetric
representations of fermionic algebras may be found in ref.~\cite{R8}.) In
addition, Jones-Smith and Mathur showed that free noninteracting
${\cal CPT}$-symmetric Dirac equations have the remarkable feature that {\it massless}
neutrinos can exhibit species oscillations~\cite{R9}.

The discovery of neutrino oscillations and the observation of the baryon
asymmetry of the universe (BAU) (see ref.~\cite{R10}) have been driving forces
in the study of the neutrino sector of the SM. Neutrino oscillations consistent
with experimental observations can occur if the SM neutrinos have small but
finite masses. The misalignment of the mass and flavour eigenbases then gives
rise to the PMNS~\cite{R11a,R11b} mixing matrix, analogous to the CKM~\cite{R12a,R12b}
mixing matrix of the quark sector. In order to generate the BAU, it is necessary
to satisfy the Sakharov conditions~\cite{R13}: namely the presence of
out-of-equilibrium dynamics and the violation of baryon number $B$, charge
${\cal C}$, and charge-parity ${\cal CP}$. Both the CKM and PMNS matrices contain a
complex phase, which provides a source of ${\cal CP}$ violation in the SM. In the
quark sector this gives rise to the ${\cal CP}$ violation observed in  $K$-,
$D$-, $B$- and $B_s$-meson mixing (see ref.~\cite{R10}). However, the magnitude
of this ${\cal CP}$ violation is insufficient to have generated the observed BAU.
An elegant framework in which both experimental observations may be accommodated
is provided by the scenario of leptogenesis~\cite{R14} (for reviews, see
refs.~\cite{R15a,R15b,R15c,R15d}). Therein, the SM is supplemented with heavy Majorana neutrinos.
The smallness of the light neutrino masses arises by means of the see-saw
mechanism~\cite{R16a,R16b,R16c,R16d,R16e} and the baryon asymmetry through the decays of the heavy
neutrinos in the expanding early universe. By virtue of the lepton-number
$L$-violating Majorana mass terms and complex Yukawa couplings, which provide an
additional source of ${\cal C}$ and ${\cal CP}$ violation, these decays are able to
generate an initial lepton excess, which is subsequently converted to a baryon
excess via the ($B+L$)-violating electroweak-sphaleron interactions of the SM~\cite{R17}.

In this article we examine an extension of QED that involves the usual Dirac
mass term $m\overline{\psi}\psi$ and an anti-Hermitian mass term $\mu\overline{\psi}\gamma^5
\psi$. The fermion field is coupled to the photon through both vector and
axial-vector couplings. The anti-Hermitian mass term is separately ${\cal C}$ even, ${\cal P}$ odd 
and ${\cal T}$ even, and is consistent
with unitarity for $\mu^2\le m^2$. We study the gauge symmetry of this model and
show that, although gauge invariance is lost in the massive case, it is
recovered in the specific situation where the Hermitian and anti-Hermitian mass
terms have equal amplitude $\mu^2=m^2$. In this limit we find that the model
reduces to that of a \emph{massless} left- or right-chiral Weyl fermion.
Moreover, we illustrate that by choosing the ratio $\mu/m$ we may obtain an
arbitrarily small but finite mass for the fermion and give more or less
prominence to one chirality. This observation, combined with the maximal
${\cal CP}$ violation of the anti-Hermitian mass term, may be directly relevant
to neutrino physics.

The paper is organized as follows: Section~\ref{s2} begins by summarizing the
essential properties of the free non-Hermitian fermion theory studied already in
refs.~\cite{R18} and~\cite{R19}. Subsequently, the gauge interactions are
introduced and the tree-level properties of the model are described. Therein,
emphasis is given to the restoration of gauge invariance in the limit $\mu^2=
m^2$. Section~\ref{s3} presents the one-loop self-energy and vertex corrections,
the details of which are given in appendix~\ref{app}. Here, the recovery of gauge
invariance is made explicit through the expected vanishing of the longitudinal
component of the vacuum polarization. Section~\ref{s4} describes an analogous
non-Hermitian Yukawa model and discusses possible implications for the neutrino
sector of the SM. A novel mechanism for generating the light neutrino masses as
well as the presence of an additional source of ${\cal CP}$ violation is
highlighted. Concluding remarks are given in section~\ref{s5}.
 
\section{Description of the Model} 
\label{s2} 
\subsection{General description}
\label{ss21}
We begin with the free fermion non-Hermitian Lagrangian considered in
ref.~\cite{R18}: 
\begin{equation}
\label{e1}
{\cal L}_0\ =\ \overline{\psi}\left(i\slashed\partial\:-\:m\:-\:\mu\gamma^5\right)\psi\;,
\end{equation}
with $\mu^2\leq m^2$, such that the energies $\omega$ are real for all
three-momenta $\vec p$; that is,
\begin{equation}
\omega^2\ =\ \vec{p}^{\,2}\:+\:M^2\ \geq\ 0~,
\end{equation}
where
\begin{equation}
M^2\ =\ m^2\:-\:\mu^2\;.
\end{equation}
It is shown in ref.~\cite{R19} that the conserved current for this model is
\begin{equation}
\label{e3}
j^\rho\ =\ \overline{\psi}\gamma^\rho\left(1\:+\:\frac{\mu}{m}\,\gamma^5\right)\psi
\end{equation}
and that the equation of motion is obtained by taking the variation of the
action with respect to $\overline{\psi}$ for fixed $\psi$. The anti-Hermitian mass term in
eq.~\eqref{e1} is even under both charge conjugation ${\cal C}$ and time-reversal 
${\cal T}$, and odd under parity ${\cal P}$. Thus, it is odd under
${\cal CPT}$. However, this does not contradict invariance under Lorentz-symmetry, since
Hermiticity has been relaxed.

In this article we gauge this model and include both vector and axial-vector
coupling to an Abelian $U(1)$ gauge field $A_\mu$:
\begin{equation}\label{e4}
{\cal L}\ =\ -\: \frac{1}{4}\,F^{\mu\nu}F_{\mu\nu}\:+\:\overline{\psi}\left[i\slashed
\partial\:-\:\slashed A(g_V\:+\:g_A\gamma^5)\:-\:m\:-\:\mu\gamma^5\right]\psi\;,
\end{equation}
where $F_{\mu\nu}=\partial_\mu A_\nu-\partial_\nu A_\mu$. In the massless case
$m=\mu=0$ the action is invariant under the combined vector and axial gauge
transformation
\begin{subequations}
\label{e5}
\begin{align}
A_\mu\ &\longrightarrow\ A_\mu\:-\:\partial_\mu\phi\;,\\ \psi\ &\longrightarrow
\ \exp\left[i\left(g_V\:+\:g_A\gamma^5\right)\phi\,\right]\psi\;,\\ \overline{\psi}\
&\longrightarrow\ \overline{\psi}\,\exp\left[i\left(-\:g_V\:+\:g_A\gamma^5\right)\phi\,
\right]\,.
\end{align}
\end{subequations}
However, in the massive case $m\ne0$ and/or $\mu\ne0$ this gauge invariance is
lost.

The free fermion propagator of this theory is
\begin{equation}\label{e6}
iS\ =\ i\,\frac{\slashed p\:+\:m\:-\:\mu\gamma^5}{p^2\:-\:M^2\:+\:i
\varepsilon}~,
\end{equation}
where $\varepsilon=0^+$. We see immediately that eq.~\eqref{e6} has a light-like
pole for $\mu=\pm\,m$ ($M^2=0$), like that of a {\it massless} theory, with the
propagator taking the form
\begin{equation}
iS\ =\ i\,\frac{\slashed p\:+\:m\,(\mathbb{I}_4\:\mp\:\gamma^5)}{p^2\:+\:i
\varepsilon}~.
\end{equation}
The mass term in the numerator is proportional to the chiral projection
operators
\begin{equation}
P_{R(L)}\ =\ \frac{1}{2}\left(\mathbb{I}_4\:+(-)\:\gamma^5\right)\;,
\end{equation}
where $\mathbb{I}_n$ is the $n\times n$ unit matrix. Separating the right- and
left-chiral components $\psi_R=P_R\psi$ and $\psi_L=P_L\psi$ in the current
(\ref{e3}), we see that the probability density may be written as
\begin{equation}\label{e8}
\rho\ =\ \left(1+\frac{\mu}{m}\right)|\psi_{\rm R}|^2+\left(1-\frac{\mu}{m}
\right)|\psi_{\rm L}|^2~.
\end{equation}
Evidently, for $\mu=+(-)\,m$ the contribution to the probability density is
entirely from the right-(left)-handed degree of freedom. Therefore, it appears
that in the limit $\mu=+\,m$ we obtain a {\it massless right-handed} theory, and
in the limit $\mu=-\,m$ we obtain a {\it massless left-handed} theory. This
feature is the focus of this article. Moreover, in Sec.~\ref{s4}, we comment on
potential implications of this non-Hermitian theory for the neutrino sector of the SM and, 
in particular, the smallness of the light-neutrino masses.

The preceding observations suggest that it proves illustrative to consider this
theory in an explicit chiral basis. We do so in the following section and
show explicitly that invariance under the gauge transformation in eq.~\eqref{e5} is
recovered in the limit $\mu\to\pm\,m$, as we would anticipate for a theory that
appears to be effectively massless.

\subsection{Chiral basis}
\label{ss22}

In order to recast eq.~\eqref{e4} in an explicit chiral basis, we first rotate from
the Dirac basis to the Weyl basis via the orthogonal transformation
\begin{equation}\psi_W\ = \ \begin{pmatrix}\psi_L \\ \psi_R \end{pmatrix}\ =\ \frac{1}{\sqrt{2}}
\begin{pmatrix} \mathbb{I}_2 & -\,\mathbb{I}_2 \\ \mathbb{I}_2 & \mathbb{I}_2
\end{pmatrix}\psi\;.\end{equation}
We may then work directly with the two-component right- and left-chiral spinors
$\psi_{R}$ and $\psi_L$. 

In the Weyl basis the gamma matrices take the form
\begin{equation}\gamma^{\mu}_W\ =\ \begin{pmatrix} 0 & \sigma^{\mu} \\ \bar{\sigma}^{\mu} & 0
\end{pmatrix}\;,\qquad \gamma_W^5\ =\ \begin{pmatrix} -\:\mathbb{I}_2 & 0 \\ 0 &
\mathbb{I}_2\end{pmatrix}\;,\end{equation}
where $\sigma^\mu=(\sigma^0,\sigma^i)$ and $\bar{\sigma}^\mu=(\sigma^0,-\,
\sigma^i)$, and $\sigma^i$ are the Pauli matrices. To avoid a proliferation of
subscripts and superscripts, throughout this paper we suppress $SU(2)$ spinor
indices (see appendix~\ref{app}). In addition, the projection operators are given by
\begin{equation} P_L\ = \ \begin{pmatrix}\mathbb{I}_2 & 0 \\ 0 & 0\end{pmatrix}\;,\qquad P_R\
=\ \begin{pmatrix} 0 & 0 \\ 0 & \mathbb{I}_2 \end{pmatrix}\;.\end{equation}
The fermionic sector Lagrangian is then
\begin{equation}\label{e9}
\mathcal{L}_{\rm ferm}\ =\ \begin{pmatrix} \psi_L^{\dag} & \psi_R^{\dag}
\end{pmatrix}\begin{pmatrix}i\bar{\sigma}\cdot D_- & -\,m_+ \\ -\,m_- & i
\sigma\cdot D_+\end{pmatrix}\begin{pmatrix}\psi_L\\ \psi_R\end{pmatrix}\;,
\end{equation}
where
\begin{equation}
m_{\pm}\ =\ m\:\pm\:\mu\;,
\end{equation}
and the covariant derivatives are given by
\begin{equation}
D_{\pm}^{\mu}\ =\ \partial^{\mu}\:+\:ig_{\pm}A^{\mu}\;,
\end{equation}
with
\begin{equation}
g_\pm\ =\ g_V\:\pm\: g_A\;.
\end{equation}
Notice that $\gamma^5$ matrices nolonger appear explicitly in the Lagrangian
eq.~\eqref{e9}. Instead, the non-Hermitian nature of this theory is manifest in the
asymmetry between the right- and left-chiral components of the original
four-component Dirac spinor.

We may study the on-shell structure of the Lagrangian in eq.~\eqref{e9}. For the
case $\mu=+\,m$ the Lagrangian takes the form
\begin{equation}\mathcal{L}_{\rm ferm}\big|_{\mu\,=\,+\,m}\ =\ \psi^{\dag}_Li\bar{\sigma}\cdot
D_-\psi_L\:+\:\psi_R^{\dag}i\sigma\cdot D_+\psi_R\:-\:2m\psi_L^{\dag}\psi_R\;,\end{equation}
giving the following equations of motion for $\psi_R$ and $\psi_L$:
\begin{subequations}
\begin{align}
\label{e10}
\frac{\delta S}{\delta\psi^{\dag}_R}\ =\ 0\quad&\Rightarrow\quad i\sigma\cdot D_+
\psi_R\ =\ 0\;,\\
\frac{\delta S}{\delta\psi^{\dag}_L}\ =\ 0\quad&\Rightarrow\quad i\bar{\sigma}\cdot
D_-\psi_L\ =\ 2m\psi_R\;.
\end{align}
\end{subequations}
Since the left-chiral field does not appear in the equation of motion for the
right-chiral field [eq.~\eqref{e10}], we may integrate it out, giving the tree-level
on-shell Lagrangian
\begin{equation}\label{e11}
{\cal L}^{\rm tree}_{\rm on-shell}\ =\ \psi_R^{\dag}i\sigma\cdot D_{+}\psi_R\;,
\end{equation}
which describes a massless theory of right-handed Weyl fermions. This is
precisely what we saw in the probability density [eq.~\eqref{e8}]. Moreover, the
on-shell Lagrangian [eq.~\eqref{e11}] respects the full vector and axial-vector gauge
invariance [see eq.~\eqref{e5}]; that is,
\begin{equation} A_\mu\ \longrightarrow \ A_\mu\:-\:\partial_\mu\phi\;,\qquad
\psi_R\ \longrightarrow \ \exp\left(ig_{+}\phi\right)\psi_R~.\end{equation}
For the case $\mu=-\,m$ we need only make the replacements $\psi_R\leftrightarrow
\psi_L$, $\bar{\sigma}\leftrightarrow\sigma$, and $D_+\leftrightarrow D_-$ in eq.~\eqref{e11}, yielding a
massless theory of left-handed Weyl fermions. The next subsection gives a more
explicit argument to justify the restoration of gauge invariance for the
light-like case $\mu^2=m^2$.

\subsection{Hidden gauge invariance}
\label{ss23}
In this subsection we show that gauge invariance is recovered when $\mu^2=m^2$.
To do so, we construct a two-component spinor basis in which gauge invariance is
explicit.

Written in block form, where the LL (left-left) element is located in the top
left $2\times 2$ block, the mass matrix is given by
\begin{equation}\bm{M}\ =\ \begin{pmatrix} 0 & m_+ \\ m_- & 0\end{pmatrix}~,\end{equation}
having eigenvalues $\pm\,M=\pm\sqrt{m^2-\mu^2}$ and eigendirections
\begin{equation} e_{\pm}\ =\ \frac{1}{\sqrt{2}}\begin{pmatrix} \pm\, x_+\\ x_-\end{pmatrix}\;
\quad\mbox{with}\quad x_\pm\equiv\sqrt{1\pm\mu/m}\;.\end{equation}
We rotate to the mass eigenbasis but first allow for a rescaling of the left-
and right-handed components:
\begin{equation}\label{e12}
\psi_{L(R)}\ \longrightarrow\ \psi_{L(R)}'\ =\ a_{L(R)}\psi_{L(R)}\;,
\end{equation}
where $a_{L(R)}$ are to be determined later, as explained below. This leads to
the transformation
\begin{equation} \begin{pmatrix} \psi_+ \\ \psi_- \end{pmatrix}\equiv \bm{V}^{-1}
\begin{pmatrix} \psi_L \\ \psi_R\end{pmatrix}\;,\end{equation}
with
\begin{equation}
\label{e13}
\bm{V}^{-1}\ =\ \frac{1}{\sqrt{2}}\begin{pmatrix} a_L x_- & a_R x_+\: \\
-a_L x_- & a_R x_+ \end{pmatrix}\;\quad\mbox{and}\quad \bm{V}\ =\
\frac{1}{\sqrt{2}}\begin{pmatrix} 1/(a_L x_-) & -1/(a_L x_-) \\
1/(a_R x_+) & 1/(a_R x_+) \end{pmatrix}~.
\end{equation}
The Lagrangian then becomes
\begin{align}
{\cal L}_{\rm ferm}\ &= \ \begin{pmatrix}\psi_+^\dagger~\psi_-^\dagger
\end{pmatrix}\bm{V^\mathsf{T}}\begin{pmatrix} i\overline{\sigma}\cdot D_- & -\,m_+
\\ -\,m_- & i\sigma\cdot D_+\end{pmatrix}\bm{V}\begin{pmatrix} \psi_+ \nonumber\\
\psi_- \end{pmatrix}\nonumber\\ & =\ \begin{pmatrix}\psi_+^\dagger ~\psi_-^\dagger
\end{pmatrix}\begin{pmatrix} A & B \\ C & D\end{pmatrix}\begin{pmatrix}\psi_+ \\
\psi_- \end{pmatrix}~,
\end{align}
where
\begin{subequations}
\begin{align}
A\ &=\ \frac{i\sigma\cdot D_+}{2a_R^2x_+^2}\:+\:\frac{i\overline{\sigma}\cdot D_-}
{2a_L^2x_-^2}\:-\:\frac{m}{a_Ra_Lx_+x_-}~,\\
B\ &=\ \frac{i\sigma\cdot D_+}{2a_R^2x_+^2}\:-\:\frac{i\overline{\sigma}\cdot D_-}
{2a_L^2x_-^2}\:-\:\frac{\mu}{a_Ra_Lx_+x_-}~,\\
C\ &=\ \frac{i\sigma\cdot D_+}{2a_R^2x_+^2}\:-\:\frac{i\overline{\sigma}\cdot D_-}
{2a_L^2x_-^2}\:+\:\frac{\mu}{a_Ra_Lx_+x_-}~,\\
D\ &=\ \frac{i\sigma\cdot D_+}{2a_R^2x_+^2}\:+\:\frac{i\overline{\sigma}\cdot D_-}
{2a_L^2x_-^2}\:+\:\frac{m}{a_Ra_Lx_+x_-}~,
\end{align}
\end{subequations}
and the two-component spinors $\psi_+$ and $\psi_-$ are given by
\begin{equation}
\label{e14}
\psi_{\pm}\ =\ \frac{1}{\sqrt{2}}\Big(x_+\psi_R'\:\pm\:x_-\psi_L'\Big)\;.
\end{equation}

The next step is to determine the coefficients $a_{L(R)}$. To make gauge
invariance explicit in the limit $\mu\to+(-)m$, that is, $x_{+(-)}\to 0$, only
the covariant derivative $D_{+(-)}$ should remain. A reasonable choice for the
field rescaling is
\begin{equation} 
\frac{1}{2a_R^2x_+^2}\ =\ \frac{x_+^2}{a^2}\qquad \mbox{and}\qquad\frac{1}{2
a_L^2x_-^2}\ =\ \frac{x_-^2}{a^2}\;,
\end{equation}
where $a$ is an overall numerical coefficient. Thus, we have
\begin{equation} a_R\ =\ \frac{a}{\sqrt{2}x^2_+}\qquad\mbox{and}\qquad a_L\ =\ \frac{a}{\sqrt{2}x_-^2}\;,\end{equation}
and we obtain
\begin{align}
\label{e15}
a^2{\cal L}_{\rm ferm}\ & =\ \begin{pmatrix} \psi_+^\dagger ~ \psi_-^\dagger
\end{pmatrix}\begin{pmatrix} x_+^2i\sigma\cdot D_++x_-^2i\overline{\sigma}\cdot D_- &
x_+^2i\sigma\cdot D_+-x_-^2i\overline{\sigma}\cdot D_-\nonumber\\
x_+^2i\sigma\cdot D_+-x_-^2i\overline{\sigma}\cdot D_- & x_+^2i\sigma\cdot D_++x_-^2i
\overline{\sigma}\cdot D_- \end{pmatrix}\begin{pmatrix} \psi_+ \\ \psi_- \end{pmatrix}\\
& \qquad-\:2 M \begin{pmatrix}\psi_+^\dagger ~ \psi_-^\dagger\end{pmatrix}
\begin{pmatrix} 1 & \mu/m \\ -\,\mu/m & -\,1 \end{pmatrix}\begin{pmatrix}\psi_+
\\ \psi_- \end{pmatrix}~.
\end{align}
Note that the mass matrix is not diagonal, even in the mass eigenbasis, because
of the anti-Hermitian mass term controlled by $\mu$.

In the limit $\mu\to\pm\,m$ the mass term vanishes, and we are left with a
massless theory that is invariant under the gauge transformation
\begin{equation} A_\mu\ \longrightarrow\ A_\mu\:-\:\partial_\mu\phi\;,\qquad
\psi_{\pm}\ \longrightarrow\ \begin{cases}\exp\left(ig_{+}\phi\right)\psi_{\pm}
\;,\quad & \mu\ =\ +\,m,\\ \exp\left(ig_{-}\phi\right)\psi_{\pm}\;,
\quad &\mu\ =\ -\,m\;.\end{cases}\end{equation}
Moreover, from eq.~\eqref{e14}, we have
\begin{equation}\psi_{\pm}\ =\ \begin{cases} \psi_R'\;,\qquad &\mu\ =\ +\,m,\\
\pm\,\psi_L'\;,\qquad &\mu\ =\ -\,m\;,\end{cases}\end{equation}
and
\begin{equation}\mathcal{L}_{\mathrm{ferm}}\ =\ \begin{cases}\psi^{\dag}_Ri\sigma\cdot D_+
\psi_R\;,\qquad &\mu\ =\ +\,m,\\
\psi^{\dag}_Li\bar{\sigma}\cdot D_-\psi_L\;,\qquad &\mu\ =\ -\,m\;,\end{cases}\end{equation}
for massless right- and left-chiral theories, as observed in the preceding
subsections.

The coefficients $a_{L(R)}$ and the transformation in eq.~\eqref{e13} are singular
in the limit $\mu\to\pm\, m$. However, the coefficients $a_{L(R)}$ do not appear
in the final Lagrangian [eq.~\eqref{e15}], which remains finite in the limit $\mu\to
\pm\,m$. Furthermore, the functional Jacobian of the field rescaling in
eq.~\eqref{e12}, although also singular, cancels out in the normalization of $Z$
with the partition function $Z_0$ of the corresponding free theory. Thus, the
limit $\mu\to\pm\,m$ may be taken safely, as done above.

\subsection{Exceptional points}
\label{ss24}
An $N$-dimensional Hermitian matrix always has $N$ real eigenvalues and
associated with each eigenvalue is a distinct eigenvector. For non-Hermitian
matrices the situation is more elaborate. Consider, for example, the
non-Hermitian $2$-dimensional matrix
\begin{equation}
A\ =\ \begin{pmatrix} a+ib & g \\ g & a-ib\end{pmatrix},
\end{equation}
where $a$ and $b$ are real parameters and $g$ is a real coupling constant.
The eigenvalues of $A$ are $E(g)=a\pm\sqrt{g^2-b^2}$. Thus, there are two phases: a 
{\it broken} phase (for $g^2<b^2$) in which the eigenvalues are complex and an {\it
unbroken} phase (for $g^2>b^2$) in which the eigenvalues are real. At the boundary
between the phases ($g=\pm\,b$) the eigenvalues merge, and the matrix is said to be
{\it defective} because there is only one eigenvalue $E=a$ and one eigenvector instead of two:
$(i,1)$ for $g=+\,b$ and $(1,i)$ for $g=-\,b$. The point $g^2=b^2$ is called an {\it exceptional point}.

In general, at an exceptional point a pair of eigenvalues of a non-Hermitian
matrix merge, and one of the eigenvectors disappears. (It is possible for more
than two eigenvalues to merge at an exceptional point, but this is not common.)
If the exceptional point occurs when a parameter $g$ has the value $g_0$, the
eigenvalues $E(g)$ exhibit a square-root singularity at $g=g_0$.

Hermitian matrices do not have exceptional points. Nevertheless, exceptional
points play a crucial role in explaining their behavior. For example, in
conventional Hermitian quantum theory the radius of convergence of a
perturbation expansion in powers of a coupling constant is precisely the
distance to the nearest exceptional point (a square-root singularity) in the
complex-coupling-constant plane~\cite{AHO}.

In the limit $\mu=\pm\,m$ the mass matrix $\bm{M}$ is defective and, as
explained above, the transformation in eq.~\eqref{e13} becomes singular. For instance,
for $\mu=+\,m$ the mass matrix has the Jordan normal form
\begin{equation} 
\bm{M}\ =\ 2\,m \begin{pmatrix} 0 & 1 \\ 0 & 0\end{pmatrix}\;,
\end{equation}
with zero eigenvalues. In this limit we have chiral flips biased from left to
right, depleting the probability density of the left-handed component, as we saw
in eq.~\eqref{e8}. In other words, we again arrive at a massless theory of
right-handed fields. Conversely, in the limit $\mu\to-\,m$ we arrive at a
massless theory dominated by left-handed fields with chiral flips biased from
right to left.

\begin{figure}
\begin{center}
\includegraphics[scale=1]{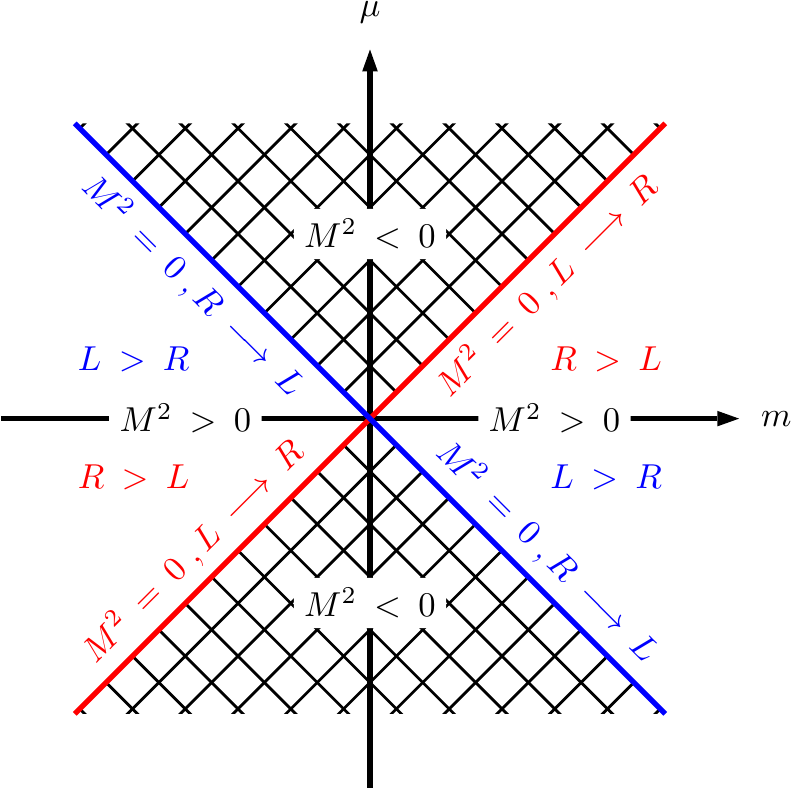}
\end{center}
\caption{\label{f1} Schematic representation of the $m$--$\mu$
plane, where the tachyonic region (cross-hatched) is bounded by the lines $\mu^2
=m^2$, along which the mass matrix becomes defective: $\mu=+\,m$ (red)
corresponds to total left-chiral domination, and $\mu=-\,m$ (blue) corresponds
to total right-chiral domination. Along the line $\mu=0$ the symmetry between
the right- and left-chiral components is restored.}
\end{figure}

The appearance of defective matrices is rare in physics, especially in field
theory, and it is worth considering what this singular behaviour signals. For
$\mu^2<m^2$ we have right- and left-chiral components with positive mass-squared
$M^2>0$ (time-like) and real-valued energies $\omega\in\mathbb{R}$. For $\mu>0$
the right-chiral component dominates; for $\mu<0$ the left-chiral component
dominates; for $\mu=0$ we have exact symmetry between both components. On the
other hand, for $\mu^2>m^2$ we still have right- and left-chiral components, but
these are now tachyonic, having negative mass-squared $M^2<0$ (space-like) and
imaginary-valued energies $i\omega\in\mathbb{R}$. For the special case $\mu^2=
m^2$ we have a massless fermion $M^2=0$ (light-like) and real-valued energies
$\omega=|\vec{p}\,|\in\mathbb{R}$. For $\mu=+\,m$ this field is completely
dominated by its right-chiral component and for $\mu=-\,m$ it is completely
dominated by its left-chiral component. The mass matrix becomes defective at the
boundary between the time-like particle and space-like tachyonic regimes. This
is indicated graphically in figure~\ref{f1}.

\section{One-loop corrections}\label{s3}

We give here the one-loop corrections to the fermion and photon self-energies,
as well as the three-point vertex. The technical details of the calculations are
given explicitly in appendix~\ref{app}. For our purposes it is convenient to express
the one-loop results in terms of the Passarino-Veltman form factors~\cite{R20},
the definitions of which are also given in appendix~\ref{app}. We work in the Feynman
gauge throughout, with the gauge-fixing term
\begin{equation} 
\mathcal{L}_{\rm gf}\ =\ \frac{1}{2}\,\big(\partial_{\mu}A^{\mu}\big)^2\;,\end{equation}
in which the photon propagator has the simple form
\begin{equation}
iD_{\mu\nu}(k)\ =\ \frac{i\eta_{\mu\nu}}{k^2+i\varepsilon}\;.
\end{equation}

\paragraph{Fermion self-energy.} There are four contributions to the one-loop fermion self-energy: one with two
vector couplings, one with two axial couplings, and two with one vector and one
axial coupling. Employing dimensional regularization and working in $d=4-2
\epsilon$ dimensions, we find the total self-energy
\begin{equation}\label{e16}
\Sigma(p)\ =\ \frac{2-d}{16\pi^2}\,(g_V-g_A\gamma^5)^2\,\slashed{p}\,B_1\:+\:
\frac{d}{16\pi^2}(g_V^2-g_A^2)(m+\mu\gamma^5) B_0\;,
\end{equation}
where we have suppressed the arguments on the form factors $B_{0;1}\equiv B_{0;
1}(p,M,0)$. Isolating the logarithmically-divergent part, we obtain
\begin{equation}\label{e17}
\Sigma(p)=\frac{1}{16\pi^2\epsilon}~\Big(\slashed p(g_V+g_A\gamma^5)^2+4(
g_V^2-g_A^2)(m+\mu\gamma^5)\Big)~+~\mbox{finite}~,
\end{equation}
where higher orders in $p$ are omitted. 

The RL and LR components of the fermion self-energy are given by
\begin{subequations}
\begin{align}
\Sigma_{RL}(p)\ &=\ \frac{g_+g_-}{16\pi^2}\,d\,m_-\,B_0\;,\\
\Sigma_{LR}(p)\ &=\ \frac{g_+g_-}{16\pi^2}\,d\,m_+\,B_0\;.
\end{align}
\end{subequations}
For $\mu=\pm\,m$ we see that one of these components vanishes such that it
remains the case that only the operator $\psi_L^\dag\psi_R$ ($\mu=+\,m$) or
$\psi_R^\dag\psi_L$ ($\mu=-\,m$) is present, preserving the argument in
Subsec.~\ref{ss22}. Specifically, the equations of motion for the right- and
left-chiral fields at order $g^2$ are given by
\begin{subequations}
\begin{align}
Z_Ri\sigma^{\mu}\cdot D_+\psi_R\ &=\ 0\;,\\
Z_Li\bar{\sigma}^{\mu}\cdot D_- \psi_L\ &=\ \left(2m\:+\:\delta m\:-\:\Sigma_{L
R}(p)\right)\psi_R\;,
\end{align}
\end{subequations}
where $\delta m$ is the mass counterterm and in the on-shell scheme the
wavefunction renormalization $Z_{R(L)}$ is given by
\begin{equation} Z_{R(L)}\ =\ 1\:+\:\frac{\mathrm{d}}{\mathrm{d}\slashed{p}}\, \Sigma_{RR(LL)}
(p)\,\bigg|_{p^2\, =\, 0}\;.\end{equation}
We may again integrate out the left-chiral component, obtaining a massless
right-handed theory also at order $g^2$.

\paragraph{Polarization tensor.} There are also four contributions to the one-loop photon polarization tensor,
and we find the total polarization tensor
\begin{equation}\label{e18}
\Pi^{\mu\nu}(p)\ = \ -\:\frac{g_V^2+g_A^2}{2\pi^2}\,\big(p^{\mu}p^{\nu}\:-\:
\eta^{\mu\nu}p^2\big)\big(B_{21}+B_1\big)\:+\:\frac{g_A^2}{\pi^2}\,\eta^{\mu\nu}
M^2B_0\;.
\end{equation}
The form factors are evaluated as $B_{0;1;21}\equiv B_{0;1;21}(p,M,M)$. As
expected from the loss of gauge invariance in the case of axially-coupled
massive fermions, the polarization tensor is not transverse and contains the
longitudinal part
\begin{equation} 
\Pi^{\mu\nu}(p)\ \supset \ \frac{g_A^2}{\pi^2}\,\eta^{\mu\nu}M^2B_0\;.\end{equation}
Nevertheless, this longitudinal part vanishes and the polarization tensor
becomes transverse when $\mu^2=m^2$ (that is, when $M=0$). Hence, as a
consequence of the restoration of gauge invariance (see Subsec.~\ref{ss23}), the
polarization tensor satisfies the standard QED Ward identity.

Isolating the logarithmic divergences in eq.~\eqref{e18}, we obtain 
\begin{equation}\Pi^{\mu\nu}(p)\ =\ \frac{g_V^2+g_A^2}{12\pi^2\epsilon}\,(p^\mu p^\nu-\eta^{
\mu\nu}p^2)+\:\frac{g_A^2}{\pi^2\epsilon}M^2\eta^{\mu\nu}\:+\:\mbox{finite}~,\end{equation}
where higher orders in $p$ are omitted.

\paragraph{Vertex.} The four different contributions to the three-point vertex lead to the total
one-loop correction 
\begin{align}
\label{e19}
&\Lambda^\mu(p,q)\: =\: \frac{2-d}{16\pi^2}(g_V+g_A\gamma^5)\nonumber\\&\qquad\qquad\qquad \times\: \Big\{(g_V+g_A
\gamma^5)^2\Big[(2-d)\gamma^\mu C_{24}+\gamma^\rho\gamma^\mu\gamma^\kappa
F_{\kappa\rho}\Big]+(g_V^2+g_A^2)\gamma^\mu M^2C_0\Big\}\nonumber\\
&\quad +\:\frac{1}{4\pi^2}\,(g_V^2+g_A^2)(m-\mu\gamma^5)\nonumber\\&\qquad\qquad\qquad \times\:\Big\{g_V\Big[p^\mu
\big(2C_{11}+C_0\big)+q^{\mu}\big(2C_{12}+C_0\big)\Big]-g_A\gamma^5(p^\mu+q^\mu)
C_0\Big\}\;,
\end{align}
where we have defined
\begin{equation} F_{\kappa\rho}\ =\ p_{\kappa}\,p_{\rho}\,\big(C_{11}+C_{21}\big)\:+\:q_\kappa
\,q_{\rho}\,\big(C_{22}+C_{12}\big)\:+\:p_\kappa\,q_{\rho}\,\big(C_{23}+C_{11}
\big)\:+\:q_{\kappa}\,p_{\rho}\,\big(C_{23}+C_{12}\big)\;.\end{equation}
The three-point form factors (see appendix~\ref{app}) are evaluated with arguments $p_1=
p$, $p_2=q$, $m_1=m_3=M$, and $m_2=0$. The divergent contribution to eq.~\eqref{e19}
arises from the form factor $C_{24}$ and is given by
\begin{equation} \Lambda^\mu\ \supset\ \frac{1}{16\pi^2\epsilon}\,(g_V+g_A\gamma^5)^3
\gamma^\mu\;,\end{equation}
which is consistent with the self-energy [eq.~\eqref{e17}], as imposed by the Ward
identity for $g_A\to0$, describing the usual gauge invariance of
vectorially-coupled massive QED.

Finally, the RL and LR components of the vertex, that is, those mediating
right-to-left and left-to-right chiral flips, are given by
\begin{subequations}
\begin{align}
\Lambda^\mu_{RL}\ &=\ \frac{g_+g_-}{4\pi^2}\,m_-\,\Big[(g_++g_-)\big(p^{\mu}\,
C_{11}+q^\mu\,C_{12}\big)\:+\:g_-(p^\mu+q^\mu)C_0\big]\;,\\[0.3em]
\Lambda^\mu_{LR}\ &=\ \frac{g_+g_-}{4\pi^2}\, m_+\,\Big[(g_++g_-) \big(p^\mu\,
C_{11}+q^\mu\,C_{12}\big)\:+\:g_+(p^\mu+q^\mu)C_0\Big]\;.
\end{align}
\end{subequations}
Like the RL and LR components of the fermion self-energies, these terms are
proportional to $m_-$ and $m_+$, respectively, so we have only left-to-right
chiral flips for $\mu=+\,m$ and right-to-left chiral flips for $\mu=-\,m$,
which preserves the structure observed in Subsec.~\ref{ss22} for $\mu^2=m^2$
also at order $g^3$.

\section{Implications for neutrino masses}\label{s4}

This section highlights potential implications of the behavior of this
non-Hermitian theory for the neutrino sector of the SM. We extend
the SM with a right-handed singlet neutrino $\nu_R$. In the Dirac
basis we write the non-Hermitian neutrino Yukawa sector, assuming only a single generation for now, as
\begin{equation} \mathcal{L}=\ \overline{L}_Li\slashed{D}L_L \:+\:\overline{\nu}_Ri\slashed{\partial}
\nu_R\:-\: h_{-}\overline{L}_L\widetilde{\phi}\nu_R\: -\:h_{+}\overline{\nu}_R
\widetilde{\phi}^{\dag}L_L\;,\end{equation}
where $L_L=(\nu_L,e_L)$ is the $SU(2)$ lepton doublet, $\widetilde{\phi}=i
\sigma_2\phi^*$ is the isospin conjugate of the Higgs doublet and $D_\mu$ is the
usual covariant derivative of the SM gauge groups. Note that we have swapped $+$
and $-$ relative to the non-Hermitian model of QED in the preceding sections.
The non-Hermitian Yukawa couplings are
\begin{equation}
h_\pm\ =\ h\:\pm\:\eta\;,
\end{equation}
where, for now, we assume that $h,\eta\in\mathbb{R}$. Since the electroweak sector of the
SM is already written in terms of chiral fields, no $\gamma^5$ appears
explicitly in the non-Hermitian Lagrangian. Even so, in the symmetry-broken
phase, the non-Hermitian Yukawa couplings give rise to a Hermitian mass $m=vh$
and an anti-Hermitian mass $\mu=v\eta$, with $m_\pm=v(h\pm\eta)$, in complete
analogy to the non-Hermitian Abelian theory considered in Sec.~\ref{s2}. 

In the unitary gauge and after spontaneous symmetry breaking, the Higgs doublet
takes the form
\begin{equation} \phi\ =\ \frac{1}{\sqrt{2}}\begin{pmatrix} 0 \\ v+H \end{pmatrix}\;,\qquad
\widetilde{\phi}\ =\ \frac{1}{\sqrt{2}}\begin{pmatrix}v+H \\ 0
\end{pmatrix}\;.\end{equation}
Hence, the neutrino sector becomes
\begin{equation}
\label{e20}
\mathcal{L}_{\mathrm{\nu}}\ =\ \overline{\nu}_{L}i\slashed{\partial}\nu_{L}
\:+\:\overline{\nu}_{R}i\slashed{\partial}\nu_{R}\:-\: h_-\frac{v}{\sqrt{2}}\,\overline{\nu}_{L}\nu_{R}\:
-\:h_+\frac{v}{\sqrt{2}}\,\overline{\nu}_{R}\nu_{L}\:-\: \frac{h_{-}}{\sqrt{2}}
\,\overline{\nu}_{L}H\nu_{R}\:
-\:\frac{h_{+}}{\sqrt{2}}\,\overline{\nu}_{R}H\nu_{L}\;.
\end{equation}
The first four terms of the Lagrangian in eq.~\eqref{e20} can be written in the matrix form
\begin{equation} 
\mathcal{L}_{\mathrm{\nu}}\ \supset\ \begin{pmatrix}\overline{\nu}_{\mathrm{L}} &
\overline{\nu}_{\mathrm{R}} \end{pmatrix} \begin{pmatrix} i\slashed{\partial} & -\, 
h_-\frac{v}{\sqrt{2}}\\ -\, h_+\frac{v}{\sqrt{2}} & i\slashed{\partial}\end{pmatrix}
\begin{pmatrix} \nu_{\mathrm{L}} \\ \nu_{\mathrm{R}} \end{pmatrix}\;,\end{equation}
where the neutrino mass matrix
\begin{equation} 
\bm{M}\ =\ \frac{v}{\sqrt{2}}\begin{pmatrix} 0 & h_{-}\\ h_{+} & 0
\end{pmatrix}
\end{equation}
has eigenvalues
\begin{equation} 
\pm\:M\ =\ \pm\,\frac{v}{\sqrt{2}}\,\sqrt{h^2-\eta^2}\;.
\end{equation}
Proceeding in analogy to Subsec.~\ref{ss23}, we make the field redefinition ($a=\sqrt{2}$)
\begin{equation} 
\nu_{L(R)}\ \longrightarrow \ \nu_{L(R)}'\ =\ \frac{\nu_{L(R)}}{x_{-(+)}^{2}}
\,\;,
\end{equation}
where
\begin{equation}
x_{\pm}\ \equiv\ \sqrt{1\pm\eta/h}\;.
\end{equation}
We then move to the mass eigenbasis spanned by the two-component spinors
\begin{equation} 
\begin{pmatrix} \nu_+ \\ \nu_- \end{pmatrix}=\frac{1}{\sqrt{2}}\begin{pmatrix}
x_+ & x_-\: \\[0.5em] x_+ & -\,x_- \end{pmatrix}\begin{pmatrix} \nu_L' \\
\nu_R'\end{pmatrix} \;,
\end{equation}
with
\begin{equation} 
\nu_{\pm}\ =\ \frac{1}{\sqrt{2}}\Big(x_+\nu_L'\:\pm\:x_-\nu_R'\Big)\;.\end{equation}
Thus, in the limit $\eta\to h$ and in analogy to Subsec.~\ref{ss23}, we obtain a
theory of massless left-handed neutrinos, which is the ``original'' Standard
Model. However, arranging for $\eta\sim h$ with $\eta<0$, we obtain a nonzero
but arbitrarily small mass for the neutrinos, with the propagating state still
dominated by its left-chiral component.

In the above minimal extension of the SM the singlet neutrino $\nu_R$ does not couple to the $SU(2)_L$ gauge fields, 
and we cannot make use of an analogy to the non-Hermitian Abelian gauge couplings of Subsec.~\ref{ss21}. However, in 
the so-called left-right SM~\cite{RLRa,RLRb,RLRc}, where the SM gauge groups are extended from $SU(2)_L\otimes U(1)_Y$ to 
$SU(2)_L\otimes SU(2)_R\otimes U(1)_{B-L}$, the $SU(2)_R$ gauge fields couple directly to the right-handed neutrino current. 
Thus, by introducing couplings $g_+=g_V+g_A$ and $g_-=g_V-g_A$ of the left- and right-handed currents to the charged gauge 
fields $W_L^{\pm\,,\mu}$ and $W_R^{\pm\,,\mu}$, and $g_+'=g_V'+g_A'$ and $g_-'=g_V'-g_A'$ to the neutral gauge fields 
$Z_L^{\mu}$ and $Z_R^{\mu}$, those of the right-handed neutrino may be suppressed for $g_V{}^{(}{}'{}^{)}\sim g_A{}^{(}{}'{}^{)}$. 
This, of course, amounts only to choosing different values for the tree-level $SU(2)_L$ and $SU(2)_R$ gauge couplings, which need 
not result from a non-Hermitian theory. Nevertheless, this construction might provide a common origin 
for such a structure in the gauge and Yukawa sectors.

The masses of the left- and right-handed neutrinos are degenerate in this
construction, both being light for $\eta\sim h$. However, since the right-handed
neutrino is still a singlet of the SM gauge groups, we are not precluded from adding a Majorana mass term
\begin{equation} 
\mathcal{L}_{\nu}\ \supset\ -\:m_{R}\,\overline{\nu}_R^{\cal C}\,\nu_R\;,
\end{equation}
where ${\cal C}$ denotes charge conjugation. In this case, the Lagrangian takes the form
\begin{equation}
-\:\mathcal{L}_{\nu}\ \supset \ \frac{1}{2}\,\begin{pmatrix} \overline{\nu}_L & \overline{\nu}_R^{\cal C} \end{pmatrix}
\begin{pmatrix} 0 & m_- \\ m_+ & m_R\end{pmatrix}\begin{pmatrix} \nu_L^{\cal C} \\ \nu_R\end{pmatrix}\:+\:
\frac{1}{2}\,\begin{pmatrix} \overline{\nu}_L^{\cal C} & \overline{\nu}_R \end{pmatrix}\begin{pmatrix} 0 & m_- \\ m_+ & m_R\end{pmatrix}\begin{pmatrix} \nu_L \\ \nu_R^{\cal C}\end{pmatrix}
\end{equation}
For $m_R\gg 2M$ the masses of the light and heavy neutrinos are $m_L=-\,M^2/m_R$ and $m_R$, which
drives up the mass of the right-handed neutrino and further suppresses that of
the left-handed neutrino by means of the see-saw
mechanism~\cite{R16a,R16b,R16c,R16d,R16e}.

It is worth commenting on the generalization to complex Hermitian and anti-Hermitian Yukawa couplings $h$ and $\eta$. In this case, the Yukawa sector takes the form
\begin{equation}
\mathcal{L}_{\mathrm{\nu}}\ \supset\ -\: h_-\frac{v}{\sqrt{2}}\,\overline{\nu}_{L}\nu_{R}\:
-\:h_+^*\frac{v}{\sqrt{2}}\,\overline{\nu}_{R}\nu_{L}\;.
\end{equation}
The mass-squared is then given by
\begin{equation}
M^2\ =\ \frac{v^2}{4}\Big(|h|^2\:-\:|\eta|^2\:-\:2i\mathrm{Im}\,h^*\eta\Big)\;,
\end{equation}
which delivers real masses only when $\mathrm{Im}\,h^*\eta$ vanishes, i.e.~when $h=\eta$. Thus, if we want small but finite masses, we are required to take $h$ and $\eta$ to be real.\

The situation is somewhat different, however, when we consider the extension of the above model to include $N$ generations: \begin{equation} 
\mathcal{L}\ =\ \overline{L}_{L}^ki\slashed{D}L_{L,k}\:+\:
\overline{\nu}_{R}^{\alpha}i\slashed{\partial}\nu_{R,\alpha}\:-\: [h_{-}]_k^{\phantom
{l}\alpha}\overline{L}_{L}^k\widetilde{\phi}\nu_{\mathrm{R},\alpha}\:-\:
[h_{+}]^k_{\phantom{k}\alpha}\overline{\nu}_{R}^{\alpha}\widetilde{\phi}^\dag
L_{L,k}\;,
\end{equation}
where we have assumed only Dirac masses in the first instance. We have employed the flavour-covariant notation of ref.~\cite{R21},
where the left- and right-handed sectors transform in the fundamental representation of two flavour groups $U_L(N)$ and $U_R(N)$, 
respectively, and flavour indices are raised and lowered by complex conjugation. We have taken the number of left- and right-handed 
fields to be equal for simplicity in what follows; this need not be the case in general. Under a general transformation in $U_L(N)\times U_R(N)$, we have
\begin{subequations}
\begin{gather}
L_{L,k} \ \longrightarrow\ L_{L,k}'\ =\ V_k^{\phantom{k}l}L_{L,l}\;,\qquad L^k_{L}\ \equiv\ (L_{L,k})^{\dag}\ \longrightarrow\ {L'_L}^k\ =\ V^k_{\phantom{k}\,l}L_L^l\;,\\
\nu_{R,\alpha}\ \longrightarrow\ \nu_{R,\alpha}'\ = \ U_{\alpha}^{\phantom{\alpha}\beta}\nu_{R,\beta}\;,
\qquad \nu_{R}^{\alpha}\ \equiv\ (\nu_{R,\alpha})^{\dag}\ \longrightarrow\ {\nu'_{R}}^{\alpha}\ = \ U^{\alpha}_{\phantom{\alpha}\beta}\nu_{R}^{\beta}\;,
\end{gather}
\end{subequations}
where $V^k_{\phantom{k}l}\equiv (V_{k}^{\phantom{k}l})^*\in U_{L}(N)$ and $U^{\alpha}_{\phantom{\alpha}\beta}\equiv (U_{\alpha}^{\phantom{\alpha}\beta})^*\in U_{R}(N)$. 
The Yukawa coupling matrices $\bm{h}_{\pm}=\bm{h}\pm\bm{\eta}$ transform as tensors of $U_L(N)\times U_R(N)$ and flavour covariance of the Lagrangian requires the transformation property
\begin{equation}
[h_{\pm}]_k^{\phantom{k}\alpha}\ \longrightarrow\ [h'_{\pm}]_{k}^{\phantom{k}\alpha}\ =\ V_{k}^{\phantom{k}l}U^{\alpha}_{\phantom{\alpha}\beta}
[h_{\pm}]_l^{\phantom{l}\beta}\;.
\end{equation}
In general, there will not exist a flavour basis in which the Yukawa matrices
$\bm{h}_+$ and $\bm{h}_-$ are simultaneously diagonal. As a result, there can be
a three-fold misalignment for general Yukawa matrices, i.e. the weak, $+$ Yukawa
and $-$ Yukawa bases can point in three different directions in flavour space.
Hence, for three generations, neutrino oscillations in this model are governed
by 6 rather than 3 mixing angles and 2 rather than 1 ${\cal CP}$-violating
phases. This additional source of ${\cal CP}$ violation is of particular
relevance to the potential embedding of this non-Hermitian theory within the
scenario of leptogenesis.

In the symmetry-broken phase, for the case of two generations ($N=2$), the mass
spectrum contains four mass eigenstates with masses given by the roots of
\begin{equation}
M_{1(2)}^2\ =\ \frac{v^2}{4}\,\Big[\mathrm{tr}\,\bm{h}^{\dag}_+\bm{h}_-\:-(+)\:
\Big(2\,\mathrm{tr}\,\big(\bm{h}^{\dag}_+\bm{h}_-\big)^2-\big(\mathrm{tr}\,
\bm{h}^{\dag}_+\bm{h}_-\big)^2\Big)^{1/2}\Big]\;.
\end{equation}
It is clear that one may obtain the massless limit by choosing $\bm{h}=\pm\,
\bm{\eta}$. However, such a constraint is not a necessary condition for
obtaining a spectrum with massless states. In the case that
\begin{equation}
\label{cond1}
\mathrm{det}\,\bm{h}^{\dag}_+\bm{h}_-\ =\ 0\qquad\Rightarrow\ \qquad\mathrm{tr}
\,\big(\bm{h}^{\dag}_+\bm{h}_-\big)^2\ =\ \big(\mathrm{tr}\,\bm{h}^{\dag}_+
\bm{h}_-\big)^2\;,
\end{equation}
we obtain two massless states ($\pm\,M_1=0$) and two states with masses given by
the roots of
\begin{equation}
M_2^2\ =\ \frac{v^2}{2}\,\mathrm{tr}\,\bm{h}^{\dag}_+\bm{h}_-\ =\ \frac{v^2}{2}
\Big[\mathrm{tr}\,\bm{h}^{\dag}\bm{h}\:-\:\mathrm{tr}\,\bm{\eta}^{\dag}\bm{\eta}
\:-\:2i\mathrm{Im}\,\mathrm{tr}\,\bm{h}^{\dag}\bm{\eta}\Big]\;.
\end{equation}
For $M_2$ to be real, we require
\begin{equation}
\label{cond2}
\mathrm{Im}\,\mathrm{tr}\,\bm{h}^{\dag}\bm{\eta}\ =\ 0\;.
\end{equation}
Subsequently imposing the additional constraint that
\begin{equation}
\label{cond3}
\mathrm{tr}\,\bm{h}^{\dag}\bm{h}\ =\ \mathrm{tr}\,\bm{\eta}^{\dag}\bm{\eta}\;,
\end{equation}
we also obtain $M_2=0$, giving a massless spectrum. In complete analogy to the
single-flavour case, we may obtain an arbitrarily small but finite mass
splitting $\Delta M^2=M_2^2-M_1^2$ by choosing
\begin{equation}
\mathrm{tr}\,\bm{h}^{\dag}\bm{h}\ \sim\  \mathrm{tr}\,\bm{\eta}^{\dag}\bm{\eta}
\;.
\end{equation}
As a result, there is the potential to obtain sub-eV scale Dirac neutrino masses
from Hermitian and anti-Hermitian Yukawa couplings, whose orders of magnitude
may themselves be much closer to the other SM Yukawa couplings and larger than
the unnatural $10^{-12}$ that would otherwise be required for agreement with
neutrino oscillation data.

Note that eqs.~\eqref{cond1}, \eqref{cond2} and \eqref{cond3} comprise three constraints
on the total of 16 parameters in the complex-valued $2\times 2$ matrices $\bm{
h}$ and $\bm{\eta}$. These three necessary conditions provide a much weaker
constraint on the elements of $\bm{h}$ and $\bm{\eta}$ than the condition
$\bm{h}=\bm{\eta}$. Moreover, they do not, as in the single-flavour case,
require $\bm{h}$ and $\bm{\eta}$ to be real-valued matrices.

As for the single-flavour case, we can include a Majorana mass term of the form
\begin{equation} 
-\:\mathcal{L}_{\mathrm{\nu}}\ \supset\ \frac{1}{2}\:\overline{\nu}_{R,\alpha}^{\cal C}m^{\alpha\beta}_{R}
\nu_{R,\beta}\:+\:\mathrm{H.c.}\;,
\end{equation}
where the mass matrix $\bm{m}_R$ transforms as a rank-2 tensor of $U_R(N)$, i.e.
\begin{equation}
m_R^{\alpha\beta}\ \longrightarrow\ m_R'{}^{\alpha\beta}\ = \ U^{\alpha}_{\phantom{\alpha}\gamma}U^{\beta}_{\phantom{\beta}
\delta}m_R^{\gamma\delta}\;. 
\end{equation}
In block form the mass terms are given by
\begin{align}
-\:\mathcal{L}_{\nu}\ &\supset \ \frac{1}{2}\,\begin{pmatrix} \overline{\nu}_L^k & \overline{\nu}_{R,\alpha}^{\cal C} \end{pmatrix}
\begin{pmatrix} 0 & [m_-]_{k}^{\phantom{k}\beta} \\ [m_+]^{\alpha}_{\phantom{\alpha}l} & m^{\alpha\beta}_R\end{pmatrix}
\begin{pmatrix} \nu_L^{{\cal C},l} \\ \nu_{R,\beta}\end{pmatrix}\nonumber\\&\qquad \qquad +\:\frac{1}{2}\,
\begin{pmatrix} \overline{\nu}_{L,k}^{\cal C} & \overline{\nu}_R^{\alpha} \end{pmatrix}\begin{pmatrix} 0 & [m_-]_{\beta}^{\phantom{\beta}k} \\ 
[m_+]^{l}_{\phantom{k}\alpha} & m_{R,\alpha\beta}\end{pmatrix}\begin{pmatrix} \nu_{L,l} \\ \nu_R^{{\cal C},\beta}\end{pmatrix}\;.
\end{align}
The mass matrix
\begin{equation}
\bm{M}\ = \ \begin{pmatrix} \bm{0} & \bm{m}_- \\ \bm{m}^{\mathsf{T}}_+ & \bm{m}_R\end{pmatrix}
\end{equation}
can be block diagonalized by a unitary transformation of the form $\widehat{\bm{M}}\ 
=\ \bm{W}^{\mathsf{T}}\bm{M} \bm{W}$, giving the physical neutrinos
\begin{equation}
\begin{pmatrix} N_{L} \\ N_{R}^{\cal C}\end{pmatrix} \ =\ \bm{W}^{\mathsf{T}}
\begin{pmatrix} \nu_L \\ \nu_R^{\cal C}\end{pmatrix}\;,\qquad \begin{pmatrix} N_{L}^{\cal C} \\ N_{R}\end{pmatrix} \ =
\ \bm{W}^{\dag}\begin{pmatrix} \nu_L^{\cal C} \\ \nu_R\end{pmatrix}\;,
\end{equation}
where the $N_{L}$ are the light neutrinos, whose mass matrix is given by the non-Hermitian see-saw formula
\begin{equation}
\bm{m}_L\ =\ -\,\bm{m}_-\bm{m}_R^{-1}\bm{m}_+^{\mathsf{T}}\;,
\end{equation}
and the $N_R$ are the heavy Majorana neutrinos, whose mass matrix is $\bm{m}_R$.

For $N=2$ the mass spectrum of the light neutrinos is given by
\begin{equation}
M_{1(2)}\ =\ -\,\frac{v^2}{4}\Big[\mathrm{tr}\,\bm{h}_-\bm{m}_R^{-1}\bm{h}_+^{\mathsf{T}}\:-(+)\:\Big(2\,\mathrm{tr}\,
\big(\bm{h}_-\bm{m}_R^{-1}\bm{h}_+^{\mathsf{T}}\big)^2\:-\:\big(\mathrm{tr}\,
\bm{h}_-\bm{m}_R^{-1}\bm{h}_+^{\mathsf{T}}\big)^2\Big)^{1/2}\Big]\;.
\end{equation}
We trivially obtain a massless spectrum for $\bm{h}=\pm\,\bm{\eta}$. However, as before, when
\begin{equation}
\label{condseesaw1}
\mathrm{det}\,\bm{h}_-\bm{m}_R^{-1}\bm{h}_+^{\mathsf{T}}\ =\ 0\qquad \Rightarrow\qquad \mathrm{tr}\,
\big(\bm{h}_-\bm{m}_R^{-1}\bm{h}_+^{\mathsf{T}}\big)^2\ =\ \big(\mathrm{tr}\,\bm{h}_-\bm{m}_R^{-1}\bm{h}_+^{\mathsf{T}}\big)^2\;,
\end{equation}
we obtain the spectrum
\begin{equation}
M_1\ =\ 0\;,\qquad M_2\ =\ -\:\frac{v^2}{2}\,\mathrm{tr}\,\bm{h}_-\bm{m}_R^{-1}\bm{h}_+^{\mathsf{T}}\;.
\end{equation}
For $M_2$ to be real, we now require
\begin{equation}
\label{condseesaw2}
\mathrm{Im}\,\mathrm{tr}\,\bm{h}_-\bm{m}_R^{-1}\bm{h}_+^{\mathsf{T}}\ =\ 0\;,
\end{equation}
and we obtain a completely massless spectrum if, in addition, we require that
\begin{equation}
\label{condseesaw3}
\mathrm{Re}\,\mathrm{tr}\,\bm{h}_-\bm{m}_R^{-1}\bm{h}_+^{\mathsf{T}}\ =\ 0\;.
\end{equation}
Again, the conditions eqs.~\eqref{condseesaw1}, \eqref{condseesaw2} and \eqref{condseesaw3} 
provide much weaker constraints on the form of the Yukawa matrices than $\bm{h}=\pm\,\bm{\eta}$. 
In addition, we can obtain an arbitrarily small but finite mass splitting $\Delta M^2$, 
independent of the Majorana mass term $\bm{m}_R$, by choosing the Yukawa couplings such that
\begin{equation}
\mathrm{Re}\,\mathrm{tr}\,\bm{h}\bm{m}_{R}^{-1}\bm{h}^{\mathsf{T}}\ \sim\ \mathrm{Re}\,\mathrm{tr}\,
\Big(\bm{\eta}\bm{m}_{R}^{-1}\bm{\eta}^{\mathsf{T}}\:
+\:\bm{h}\bm{m}_{R}^{-1}\bm{\eta}^{\mathsf{T}}\:-\:\bm{\eta}\bm{m}_{R}^{-1}\bm{h}^{\mathsf{T}}\Big)\;.
\end{equation}
This ability to tune the mass splitting of the light neutrinos independent of the magnitude of the 
Majorana mass term may have interesting implications 
in the light of the combined constraints provided by neutrino oscillation data and the current limits 
on lepton-flavour-violating and lepton-number-violating 
observables, including neutrinoless double-beta decay.

A comprehensive phenomenological study of the aforementioned variations of this non-Hermitian 
Yukawa model in the context of current constraints from collider 
experiments and both astrophysical and cosmological observations (for recent reviews, see refs.~\cite{R22a,R22b}) 
is beyond the scope of this article and will be presented elsewhere.

\section{Conclusions}\label{s5}

We have considered an extension of QED, whose non-Hermitian nature permits the
symmetry between the left- and right-chiral components of a Dirac fermion to be
broken by the presence of an anti-Hermitian mass term. We have shown that the
full gauge invariance of this theory is restored when the Hermitian and
anti-Hermitian masses are of equal magnitude. Moreover, we have highlighted an
intriguing possibility for explaining the smallness of the light neutrino masses
and providing an additional source of ${\cal CP}$ violation through an analogous 
extension of the SM. Further phenomenological studies
of this model and its variations are required in the context of the current
low-energy neutrino data as well as both cosmological and astrophysical
observations. 

Finally, we mention another direction of study, which deals with the dynamical
generation of the non-Hermitian mass term through nonperturbative quantum
effects. Dynamical mass generation for neutrinos (with a vanishing bare mass)
has been obtained in the context of Lorentz-symmetry violation~\cite{R23a,R23b,R23c}, where
the physical mass scale is provided by higher-order spatial derivatives. A
nonperturbative mechanism could also be responsible for the non-Hermitian mass
term in the present context, although the natural mass scale would be provided
by the Higgs mechanism, instead of Lorentz-symmetry-violating operators. In
order to explore this avenue, one needs to derive a nonperturbative gap equation
and study the possibility of a non-Hermitian mass term solution. Such a
nontrivial solution could arise in a theory involving an axial coupling and is
left for future work.

\acknowledgments
\noindent The work of P.M. is supported by a University Foundation Fellowship
(TUFF) from the Technische Universit\"at M\"unchen and by the Deutsche
Forschungsgemeinschaft cluster of excellence Origin and Structure of the
Universe.

\appendix

\section{One-loop corrections}
\label{app}

This appendix summarizes the technical details of the one-loop calculations
described in Sec.~\ref{s3}. The elements $A_{I\!J}$ of a matrix $\bm{A}$ in the
chiral field space are indexed by upper-case Roman indices $I,J,K,M,\dots=L,R$,
where the LL element is in the top left.

\paragraph{Passarino-Veltman parametrization.} In $d=4-2\epsilon$ the two-point Passarino-Veltman form factors~\cite{R20} are
\begin{equation} 
B_{0;\mu;\mu\nu}(p,m_1,m_2)\ =\ \int\!\frac{\mathrm{d}^dk}{i\pi^2}\;
\frac{1;k_{\mu};k_\mu k_\nu}{\big(k^2-m_1^2+i\varepsilon\big)\big((p+k)^2-m_2^2
+i\varepsilon\big)}~.
\end{equation}
These may be related to the scalar form factors $B_1$, $B_{21}$ and $B_{22}$ via
\begin{subequations}
\begin{align}
B_{\mu}(p,m_1,m_2)\ &=\ p_{\mu}B_1(p,m_1,m_2)\;,\\
B_{\mu\nu}(p,m_1,m_2)\ &=\ p_{\mu}p_{\nu}B_{21}(p,m_1,m_2)\:+\:\eta_{\mu\nu}
B_{22}(p,m_1,m_2)\;,
\end{align}
\end{subequations}
whose divergent parts are
\begin{gather}
B_0(p,m_1,m_2)\ \supset\ \frac{1}{\epsilon}\;,\qquad B_1(p,m_1,m_2) \ \supset\
-\:\frac{1}{2\epsilon}\;,\\
B_{21}(p,m_1,m_2)\ \supset\ \frac{1}{3
\epsilon}\;,\qquad B_{22}(p,m_1,m_2)\ \supset\ -\:\frac{1}{4\epsilon}\,
\left(m_1^2+m_2^2+\frac{p^2}{3}\right)\;.
\end{gather}
In addition, we make use of the algebraic identities
\begin{subequations}
\begin{gather}
p^2B_1(p,m_1,m_2)\ =\ \frac{1}{2}\big[A(m_1)\:-\:A(m_2)\:-\:(p^2-m_1^2-m_2^2)
B_0(p,m_1,m_2)\big]\;,\\
p^2B_{21}(p,m_1,m_2)\:+\:d B_{22}(p,m_1,m_2)\ =\ A(m_2)\:+\:m_1^2B_0(p,m_1,m_2)
\;,\\
p^2B_{21}(p,m_1,m_2)\:+\:B_{22}(p,m_1,m_2)\ =\ \frac{1}{2}\big[A(m_2)\:+\:
(m_1^2-m_2^2-p^2)B_1(p,m_1,m_2)\big]\;,
\end{gather}
\end{subequations}
where $A(m)$ is the tadpole form factor
\begin{equation}
A(m)\ =\ \int\!\frac{\mathrm{d}^dk}{i\pi^2}\;\frac{1}{k^2-m^2+i\varepsilon}
\;.
\end{equation}
Lastly, for $m_1=m_2$ we have the identity
\begin{equation}
B_1(p,m,m)\ =\ -\:\frac{1}{2}\,B_0(p,m,m)\;.
\end{equation}

The three-point form factors are
\begin{equation}
C_{0;\mu;\mu\nu}\ =\ \int\!\frac{\mathrm{d}^dk}{i\pi^2}\;\frac{1;k_\mu;k_\mu
k_\nu}{(k^2-m_1^2+i\varepsilon)\big((k+p_1)^2-m_2^2+i\varepsilon\big)\big(
(k+p_1+p_2)^2-m_3^2+i\varepsilon\big)}\;,
\end{equation}
where the arguments of $C_{0;\mu;\mu\nu}\equiv C_{0;\mu;\mu\nu}(p_1,p_2,m_1,m_2,
m_3)$ have been suppressed for notational brevity. We also define scalar form
factors via
\begin{subequations}
\begin{align}
C_\mu\ &= \ p_{1\mu}C_{11}\:+\:p_{2\mu}C_{12}\;,\\
C_{\mu\nu}\ &=\ p_{1\mu}p_{1\nu}C_{21}\:+\:p_{2\mu}p_{2\nu}C_{22}+p_{1(\mu}p_{2
\nu)}C_{23}\:+\:\eta_{\mu\nu}C_{24}\;.
\end{align}
\end{subequations}
The only divergent form factor is $C_{24}$, having the logarithmic divergence
\begin{equation}
C_{24}\ \supset \ \frac{1}{4\epsilon}~.
\end{equation}

\paragraph{Feynman rules.} In the chiral basis, the Feynman rules of the model are~\cite{R24a,R24b,R24c,R27}
\begin{itemize}
\item To each photon line associate the factor (in the Feynman gauge)
\begin{equation}
iD_{\mu\nu}(p)\ =\ \frac{i\eta_{\mu\nu}}{p^2+i\varepsilon}\;.
\end{equation}
\item To each chiral fermion line associate the factor
\begin{equation}
iS_{I\!J}(p)\ =\ i\,\frac{\delta_{I\!J}\,\bar{\sigma}_{J}\cdot p\,+M_{I\!J}}
{p^2-M^2+i\varepsilon}\;.
\end{equation}
To avoid proliferation of sub- and superscripts, the spinor index
assignment, denoted by the lower-case Gothic characters $\mathfrak{a}$ and $\mathfrak{b}$, is understood as follows:
\begin{subequations}
\begin{align}
iS_{LL}(p)\ &\equiv\ [iS_{LL}(p)]_{\mathfrak{a}\dot{\mathfrak{b}}} \ =\ \frac{ip\cdot
\sigma_{\mathfrak{a}\dot{\mathfrak{b}}}}{p^2-M^2+i\varepsilon}\;,\\
iS_{RR}(p)\ &\equiv\ [iS_{RR}(p)]^{\dot{\mathfrak{a}}\mathfrak{b}} \ =\ \frac{ip\cdot\bar{
\sigma}^{\dot{\mathfrak{a}}\mathfrak{b}}}{p^2-M^2+i\varepsilon}\;,\\
iS_{RL}(p)\ &\equiv\ [iS_{RL}(p)]^{\dot{\mathfrak{a}}}_{\phantom{\mathfrak{a}}\dot{\mathfrak{b}}} \ =\ \frac{im_-\delta^{\dot{\mathfrak{a}}}_{\phantom{\mathfrak{a}}\dot{\mathfrak{b}}}}{p^2-M^2+i\varepsilon}\;,\\
iS_{LR}(p)\ &\equiv\ [iS_{LR}(p)]_{\mathfrak{a}}^{\phantom{\mathfrak{a}}\mathfrak{b}}\ =\ \frac{
im_+\delta_{\mathfrak{a}}^{\phantom{\mathfrak{a}}\mathfrak{b}}}{p^2-M^2+i\varepsilon}\;,
\end{align}
\end{subequations}
with
\begin{equation} 
\bar{\sigma}^{\mu}_{L}\ =\ \sigma^{\mu}\ \equiv\ \sigma^{\mu}_{\mathfrak{a}\dot{\mathfrak{b}}}\;,\qquad\bar{\sigma}^{\mu}_{R}\ =\ \bar{\sigma}^{\mu}\ \equiv\
\bar{\sigma}^{\mu,\dot{\mathfrak{a}}\mathfrak{b}}\;.
\end{equation}

\item To each vertex associate a factor of $-\,ig_{I\!J}\sigma_{J}^{\mu}$, where
$\mathbf{g}=\mathrm{diag}\,(g_-,g_+)$.

\item For any closed fermion loop include a factor of $-1$ and trace over the
Lorentz indices.
\end{itemize}

In the calculation of the one-loop corrections outlined below, we also make
heavy use of the product and trace identities of the Pauli matrices, as listed
in Appendix B of~\cite{R27}.

\paragraph{Fermion self-energy.} The one-loop chiral fermion self-energies are given by
\begin{equation}
i\Sigma_{I\!J}(p)\ =\ (-\,i)^2\,g_{I\!K}\,g_{N\!J}\int\!\frac{\mathrm{d}^dk}{(
2\pi)^4}\;\sigma^\mu_K\,iS_{K\!N}(k+p)\,\sigma^\nu_N\,iD_{\mu\nu}(k)\;,
\end{equation}
where we note that the couplings $g_\pm$ are dimensionful for $d=4-2\epsilon$.
The numerator is
\begin{equation}
(2-d)\,\delta_{K\!N}\,\sigma_{N}\cdot k\:+\:d\,M_{K\!N}\,.
\end{equation}
Rewriting in terms of the Passarino-Veltman form factors, we get
\begin{equation} 
\Sigma_{I\!J}(p)\ =\ \frac{1}{16\pi^2}\,g_{I\!K}\,g_{N\!J}\big[\big(2-d)\,
\delta_{K\!N}\,\sigma_{K}\cdot p\,B_1(p,M,0)\:+\:d\,M_{K\!N}\,B_0(p,M,0)\big]\;.
\end{equation}
Hence, we obtain
\begin{subequations}
\begin{align}
\Sigma_{LL}\ &=\ \frac{g_-^2}{16\pi^2}\,(2-d)\,\bar{\sigma}\cdot p\,B_1(p,M,0)
\;,\\
\Sigma_{RR}\ &=\ \frac{g_+^2}{16\pi^2}\,(2-d)\,\sigma\cdot p\,B_1(p,M,0)\;,
\\[0.3em]
\Sigma_{RL}\ &=\ \frac{g_+g_-}{16\pi^2}\,d\,\,m_-\,B_0(p,M,0)\;,
\\[0.3em]
\Sigma_{LR}\ &=\ \frac{g_+g_-}{16\pi^2}\,d\,\,m_+\,B_0(p,M,0)\;.
\end{align}
\end{subequations}

The full fermion self-energy of the original Dirac field is obtained from the
sum over the chiral indices $I$ and $J$ with correct weighting by projection
operators. Specifically,
\begin{equation} 
\Sigma\ =\ P_R\gamma^0 \Sigma_{LL} P_L\:+\: P_L\gamma^0\Sigma_{RR} P_R\:+\:
P_L\Sigma_{RL} P_L\:+\: P_R\Sigma_{LR} P_R\;,
\end{equation}
giving
\begin{equation}
\Sigma\ =\ \begin{pmatrix} 0 & 0 \\ 1 & 0\end{pmatrix}\Sigma_{LL}\:+\:
\begin{pmatrix}0&1\\ 0 & 0\end{pmatrix}\Sigma_{RR}\:+\:\begin{pmatrix}1&0\\ 0
& 0\end{pmatrix}\Sigma_{RL}\:+\:\begin{pmatrix} 0 & 0 \\ 0 & 1\end{pmatrix}
\Sigma_{LR}\;,
\end{equation}
Summing these contributions, we obtain the result in eq.~\eqref{e16}.

\paragraph{Polarization tensor.} The polarization tensor is given by
\begin{equation}
i\Pi^{\mu\nu}(p)\ =\ (-\,1)(-\,i)^2\,g_{I\!K}\,g_{N\!J}\,\mathrm{Tr}\int\!
\frac{\mathrm{d}^dk}{(2\pi)^4}\;\sigma_K^\mu\,iS_{I\!J}(p+k)\,\sigma_N^\nu\,
iS_{N\!K}(k)~.
\end{equation}
Performing the trace over the Lorentz indices, we obtain the numerator
\begin{equation} 
2\big[\big(2k^{\mu}k^{\nu}\:+\:p^{(\mu}k^{\nu)}\:-\:\eta^{\mu\nu}k^2\:-\:
\eta^{\mu\nu}p\cdot k\big)\delta_{I\!J}\,\delta_{N\!K}\:+\:\eta^{\mu\nu}\,M_{I
\!J}\,M_{N\!K}\:+\:i\eta_{I\!J\!N\!K}\, \varepsilon^{\mu\kappa\nu\lambda}\,
(p+k)_{\kappa}k_{\lambda}\big]\;,
\end{equation}
where $\varepsilon^{\mu\kappa\nu\lambda}$ is the Levi-Civita tensor. Here, we
have defined $\eta_{I\!J\!N\!K}=1$, if $I=J=K=N=L$, $\eta_{I\!J\!N\!K}=-1$, if
$I=J=N=K=R$, and $\eta_{I\!J\!N\!K}=0$ otherwise.

Rewriting in terms of the Passarino-Veltman form factors, we are left with
\begin{align}
\Pi^{\mu\nu}(p)\ &= \ -\:\frac{g_{I\!K}\,g_{N\!J}}{4\pi^2}\,\Big\{\big(p^{\mu}p^\nu
-\eta^{\mu\nu}p^2\big)\big(B_{21}+ B_1\big)\delta_{I\!J}\delta_{N\!K}\nonumber\\&\qquad -\:
\eta^{\mu\nu}\big(M^2\delta_{I\!J}\delta_{N\!K}-M_{I\!J}M_{N\!K}\big)B_0\nonumber\\
& \qquad \qquad +\:i\varepsilon^{\mu\kappa\nu\lambda}\Big[p_{\kappa}p_{\lambda}
\big(B_{21}+B_1\big)\:+\: \eta_{\kappa\lambda}\, p^2 B_{22}\Big]\eta_{I\!J\!N\!
K}\Big\}\;.
\end{align}
When we sum over the chiral indices, the terms proportional to the Levi-Civita
tensor cancel, and we obtain the result in eq.~\eqref{e18}.

\paragraph{Three-point vertex.} The three-point vertices are given by
\begin{equation}
i\Lambda^\mu_{I\!J}(p,q)\ =\ (-i)^3\,g_{I\!K}\,g_{N\!P}\,g_{QJ}\int\!\frac{
\mathrm{d}^dk}{(2\pi)^4}\;\sigma_{K}^\nu\,iS_{K\!N}(k)\,iD_{\nu\lambda}(k+p)\,
\sigma_{P}^\mu\,iS_{PQ}(k+p+q)\,\sigma_{Q}^\lambda\;,
\end{equation}
where $p$ and $q$ are the fermion momenta. The numerator is proportional to
\begin{align}
&(2-d)\,\Big[\,\sigma^{\rho}_Q\,\bar{\sigma}_P^{\mu}\,\sigma_K^{\kappa}\,k_{
\kappa}\,(k+p+q)_{\rho}\,\delta_{K\!N}\,\delta_{PQ}\:+\:\sigma_P^{\mu}\,M_{K\!N}
\,M_{PQ}\,\Big]\nonumber\\ &\qquad +\:4\,\Big[\,k^{\mu}\,\delta_{K\!N}\,M_{PQ}\:+\:
(k+p+q)^\mu\,M_{K\!N}\,\delta_{PQ}\,\Big]\;,
\end{align}
such that the vertices can be written
\begin{align}
\Lambda^\mu_{I\!J}\ &=\ \frac{g_{I\!K}\,g_{N\!P}\,g_{QJ}}{16\pi^2}\nonumber\\&\quad \times\:\Big\{(2-d)
\Big[\Big((2-d)\,\sigma_P^{\mu}\,C_{24}\:+\:\sigma_Q^{\rho}\,\bar{\sigma}_P^{
\mu}\,\sigma_K^{\kappa}\,F_{\kappa\rho}\Big)\,\delta_{K\!N}\,\delta_{PQ}\:+\:\,
\sigma^{\mu}_P\,M_{K\!N}\,M_{PQ}\Big]\nonumber\\ &
\quad\quad +\:4\,\big(p^{\mu}\,C_{11}\:+\:q^{\mu}\,C_{12}\big)\big(\delta_{K\!N}\,
M_{PQ}\:+\:M_{K\!N}\,\delta_{PQ}\big)\:+\:4\,(p+q)^{\mu}\,C_0\,M_{KN}\,\delta_{
PQ}\,\Big\}\:.
\end{align}
Herein, we have defined
\begin{equation}
F_{\kappa\rho}\ =\ p_{\kappa}\,p_{\rho}\,\big(C_{11}+C_{21}\big)\:+\:q_{\kappa
}\,q_{\rho}\,\big(C_{22}+C_{12}\big)\:+\:p_{\kappa}\,q_{\rho}\,\big(C_{23}+
C_{11}\big)\:+\:q_{\kappa}\,p_{\rho}\,\big(C_{23}+C_{12}\big)\;.
\end{equation}
The three-point form factors are evaluated at $p_1=p$, $p_2=q$, $m_1=m_3=M$ and
$m_2=0$. Hence, element by element, we find that
\begin{subequations}
\begin{align}
\Lambda^\mu_{LL}\ &=\ \frac{g_-^2}{16\pi^2}\,(2-d)\Big[g_-\Big((2-d)\bar{
\sigma}^\mu C_{24}\:+\:\bar{\sigma}^\rho\sigma^\mu\bar{\sigma}^\kappa F_{\kappa
\rho}\Big)\:+\:g_+\bar{\sigma}^{\mu}M^2C_0\Big]\;,\\
\Lambda^\mu_{RR}\ &=\ \frac{g_+^2}{16\pi^2}\,(2-d)\Big[g_+\Big((2-d)\sigma^\mu
C_{24}\:+\:\sigma^\rho\bar{\sigma}^\mu\sigma^\kappa F_{\kappa\rho}\Big)\:+\:g_-
\sigma^\mu M^2C_0\Big]\;,\\[0.3em]
\Lambda^\mu_{RL}\ &=\ \frac{g_+g_-}{4\pi^2}\,m_-\,\Big[(g_++g_-)\big(p^{\mu}\,
C_{11}+q^\mu\,C_{12}\big)\:+\:g_-(p^{\mu}+q^\mu)C_0\big]\;,\\[0.3em]
\Lambda^\mu_{LR}\ &=\ \frac{g_+g_-}{4\pi^2}\, m_+\,\Big[(g_++g_-) \big(p^\mu\,
C_{11}+q^{\mu}\,C_{12}\big)\:+\:g_+(p^{\mu}+q^\mu)C_0\Big]\;.
\end{align}
\end{subequations}
As in the case of the self-energy, we have
\begin{equation} 
\Lambda^\mu\ =\ \begin{pmatrix} 0 & 0 \\ 1 & 0\end{pmatrix}\Lambda^{\mu}_{LL}
\: +\:\begin{pmatrix} 0 & 1 \\ 0 & 0\end{pmatrix}\Lambda^{\mu}_{RR}\:+\:
\begin{pmatrix} 1 & 0 \\ 0 & 0\end{pmatrix}\Lambda^{\mu}_{RL}\:+\:
\begin{pmatrix} 0 & 0 \\ 0 & 1\end{pmatrix}\Lambda^{\mu}_{LR}\;.
\end{equation}
Summing over the contributions, we obtain the result for the total vertex in
eq.~\eqref{e19}.

\begin{small}

\end{small}
\end{document}